\documentstyle[12pt,moriond]{article}
\begin{document}
\heading{Cold gas, the HI 21cm line and evolving galactic potentials}

\author{F.H. Briggs} {Kapteyn Astronomical Institute}
{Groningen, Netherlands} 

\begin{moriondabstract}
Neutral hydrogen  traces gravitational potentials. In the nearby
universe, 21cm emission-line surveys show that
the bulk of the HI resides in well-formed, optically-luminous galaxies.
At high redshift, 21cm line absorption against background radio
quasars occurs in gas-rich systems identified with the highest
HI column densities -- the ``damped Lyman alpha''  
quasar absorption-line systems.
High spatial-resolution observations of the redshifted 21cm line absorbers 
measure sizes and kinematics of the neutral absorbers.
\end{moriondabstract}

\def\putplot#1#2#3#4#5#6#7{\begin{centering} \leavevmode
\vbox to#2{\rule{0pt}{#2}}
\includegraphics{#1}
\end{centering}}

\section{Introduction}

Neutral hydrogen in the nearby Universe is observed to reside in the
gravitational potential wells associated with galaxies. In optically
luminous galaxies, the neutral gas is a minor fraction of the
baryonic mass, while in some extreme low surface brightness
dwarfs, the mass in gas is comparable to or may even exceed the
mass in luminous stars.  The standard models for primordial
nucleosynthesis require that the total baryonic mass
density of the Universe exceed that observed in 
luminous matter (Tytler et al 1996, Tosi et al 1998), 
and it appears that the majority
of the Universe's baryons are spread through the ionized intergalactic
medium, IGM, (Burles \& Tytler 1996, Shull et al 1996, Rauch et al 1997).
The role of the galactic potentials is to confine the atomic
gas to sufficient density that it can be self-shielding in the face
of the ionizing background radiation.

Although only a minor constituent in the $z=0$ Universe, the neutral gas is
a useful kinematic tracer of the confining gravitational potentials
inhabited by galaxies, and a large research effort has
used the 21cm line to establish the existence of dark matter in
galaxies and  to begin to decode its distribution (cf. van Albada et al 
1985, and the contribution by Swaters to these proceedings).

Quasar absorption-line surveys for damped Lyman-$\alpha$ DLa lines
indicate that a greater fraction of
the Universe's baryons was neutral in the redshift interval
$z=2$ to 3 (Wolfe et al 1995, Lanzetta et al 1995, Storrie-Lombardi et al
1996). 
In fact, the mass in neutral gas exceeded the mass in
stars at that time. Absorption line studies of the low column
density clouds in the Lyman-$\alpha$ forest show that the IGM
has already been re-ionized by the epoch at which the
highest redshift background quasars are presently observed,
at $a\approx 5$ (Schneider et al 1991, Fan et al 1999).  
The lines of sight through the DLa
absorber at subsequent epochs must then be probing  
neutral gas confined to the potential wells of evolving
gravitationally bound systems. These systems may
evolve to the galaxies of today or may be the protogalactic
clumps that merge to form the present day galaxies.  
Sections 3 and 4 of this review
return to the observational tools that the 21cm line provides for
characterizing the size and kinematics in these high $z$ systems

The following section addresses the question of whether there might be
fossil relics of these original gravitationally bound systems surviving
in the nearby Universe until the present.  Could there be a
population of dark matter mini-halos, capable of capturing and
confining neutral hydrogen in a primitive state from a formation
epoch at high $z$ until the present?  If so, then these would be
convenient and highly informative objects to study in detail.

\section{The Significance of the Galactic High Velocity Clouds}

Blitz et al (1999) recently presented a case for the Galactic
population of High Velocity Clouds(HVCs) being ${\sim}10^8$M$_{\odot}$
dark matter mini-halos, each containing ${\sim}10^7$M$_{\odot}$
of neutral gas of primordial composition, unpolluted by
mass loss from stars.  Each individual
cloud must be stable on cosmic time scales,
implying that the neutral gas is a minor dynamical component in
comparison to the dissipationless dark matter, which is responsible
for the binding potential.

Since the HVCs contain no resident stellar populations, a
significant hurdle in understanding the
physical properties of the HVCs is the absence of a clear distance indicator (such
as stars that would be accessible to spectroscopic parallax methods).
Thus, one particularly attractive aspect of the Blitz et al hypothesis is
 the construction of an independent distance indicator based on the
dynamical stability of each cloud. The only adjustable parameter
in this picture is $f$, the fraction of the total dynamical mass
composed of neutral gas. Blitz et al favor $f\approx 0.1$
with the implication that the clouds are at distances of roughly
1~Mpc, forming a Local Group population rather than a Galactic Halo
population, and typically have HI masses around ${\sim}10^7$M$_{\odot}$.
At this mass level, the HVC population adds significantly to the integral
HI content of the Local Group.

Theoretical support for this interpretation of the
HVCs comes from CDM simulations (cf. Klypin et al 1999, Moore et al 1999),
which find that many of the dark matter mini-halos that are the
building blocks of galactic systems survive intact through the
mergers and accretion events that go into building a system like
our Local Group. A population of several hundred of these mini-halos
would be expected in the Local Group. This number far exceeds the
number of star bearing satellite galaxies in the Local Group 
(cf Mateo 1998) and is more consistent with the number of HVCs.

If the HVC population is an important component of the Universe
and contributes significantly to the formation of all galaxies and
groups of galaxies, then HVC populations should exist around
other galaxies in the nearby Universe in addition to the galaxies
of the Local Group.  Furthermore, if the HVCs are representative
of a wide-spread extragalactic population, they could be present
outside groups and galaxy halos as independent entities. In fact, the
neutral gas masses of $M_{HI}\geq 10^7$M$_{\odot}$ are large
enough that emission from these objects
could be detected in 21cm line observations of nearby galaxies and groups.

A number of surveys have now been conducted that are capable of
sensing the presence of an extragalactic HVC population. Weinberg 
et al 1991 were  searching specifically for such a population of
CDM mini-halos in their VLA survey but found none.
Several extragalactic HI surveys of substantially larger volumes to
more sensitive HI mass limits have also found no objects with HVC
properties (Zwaan et al 1997, Spitzak \& Scheider 1999, Kilborn 1999).

Consider the Arecibo HI Strip Survey (Zwaan et al 1997, Sorar 1994)
as an example. The survey was an unbiased probe of the HI content
of the nearby Universe, conducted without prior recognition of the
optically identified galaxies in the survey strips.
The flavor of the survey as applied to the HVC question
(Zwaan \& Briggs 1999) is presented in Fig.~1.  Sorar (1994)
conducted the original
survey by repeatedly scanning strips of constant declination 
(over the course of as many as 30 days)
with the sensitive, high resolution 3$'$ beam of the Arecibo telescope.
The thin horizontal
lines in Fig.~1 mark the survey areas, which consist of
two declination strips, each of length ${\sim}10^h$
of right ascension to a depth of 7500 km~s$^{-1}$.
Although the survey
declinations were selected at random, the strips chance to probe
the halos of many galaxies at impact parameters of $\leq 1$ Mpc.
In total, the survey probed  $\sim$200 catalogued galaxy halos 
and the environments of $\sim$14 catalogued groups with sensitivity to neutral
hydrogen masses $\geq 10^{7}$M$_{\odot}$. Zwaan and Briggs (1999)
applied the formulation of Blitz et al to the Wakker and 
van Woerden (1991) HVC catalog to deduce distribution functions for
comparable HVC populations in these extragalactic settings, leading
to the conclusion that the survey should have made $\sim$75
HVC detections in groups and $\sim$380 detections around galaxies.
No objects with the properties of HVCs were discovered. In the
context of the Blitz et al model, Zwaan and Briggs conclude
that the $f$ parameter must be adjusted
to $f\leq 0.02$, which lowers the HI masses by a factor of 25 and
positions the objects at typical distances of ${\sim}$200 kpc --
thereby assigning them to the Galactic halo, rather than giving them Local
Group membership.

 
\begin{figure}[h]
\putplot{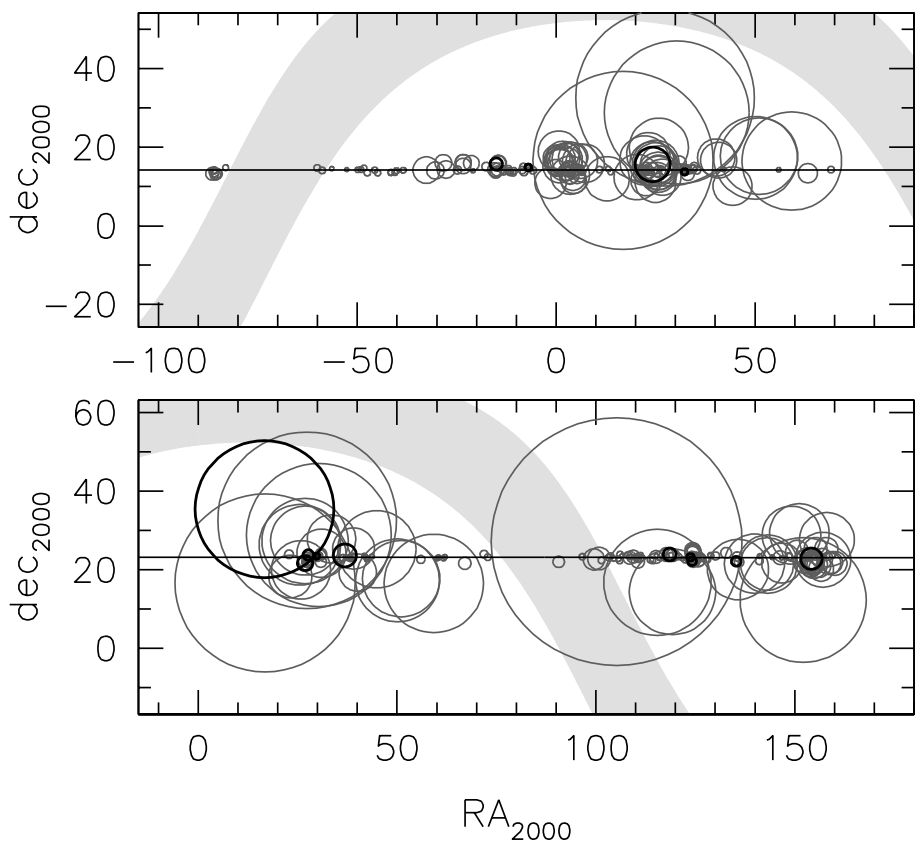}{11.5 cm}{0}{140}{140}{-130}{-560}
\caption{The galaxy groups and galaxy halos probed by the
Arecibo HI Strip Survey (Zwaan \& Briggs 1999, astro-ph/0001016).
The solid horizontal lines show the paths of the
Arecibo beam. Circles are drawn
with radii of 1 Mpc around catalogued galaxies (thin circles) and galaxy
groups (thick circles).  The shaded areas indicate the Zone of Avoidance
where $|b|<10^\circ$.  
}
\end{figure}

What did the Arecibo Survey detect?  Zwaan et al (1997) present the
follow-up and analysis for the 66 detected signals.  As with comparable surveys
conducted to date (Szomoru et al 1994, Henning 1995, Henning et al 1998,
Kraan-Korteweg et al 1998, Spitzak \& Schneider 1999, Kilborn et al 1999),
all detected signals located away from Galactic extinction and from bright
foreground stars could be optically identified with galaxies containing
stars. No surprising new populations of intergalactic clouds or 
ultra-low-surface brightness objects were uncovered.

The strong association of the neutral gas reservoirs in the nearby
Universe with stars suggests there is some causal relation. Perhaps
the confinement of HI to sufficient densities that it can remain
neutral also leads to conditions where it is difficult to avoid the
instabilities that lead to cooling, collapse and star formation. 
In such a picture, the shallower potential wells of the lower mass
dwarf galaxies would be the places where the HI is more gently 
confined and evolutionary processes would generally proceed 
more slowly. Indeed, this is the domain inhabited by the dimmest
of the gas-rich LSB galaxies.

The HVC population might form the extension of the gas-rich
dwarf population to extremely low masses, much in the spirit of
the proposal by Blitz et al.  Closing the loop between the Galactic HVCs
and the extragalactic analogs will be an important step in understanding
their nature and may form important evidence in the case for how
galaxies formed and accumulated their interstellar media.

In the meanwhile, we are left with the conclusions of Wakker and van
Woerden (1997), who after extensive review of the HVC literature, find
that no single origin can
account for the properties of the HVC population.
Instead, several mechanisms,
including infalling extragalactic clouds, cloud circulation within the
Galactic halo driven by a galactic fountain, and a warped outer arm
extension to the Galaxy must be invoked.  Explanations
for the Magellanic Stream and
associated HVC complexes require tidal
interactions within the Local Group (Putman et al 1998), and it may
be that measurement of metal abundances in the HVCs using absorption-line
techniques may be the best discriminant between ``primordial'' objects
and the remnant debris of merging.

\section{Damped Lyman-$\alpha$ absorbers and the HI content of the Lyman Break Galaxies}

Quasar absorption line studies of the Damped Lyman-$\alpha$ absorbers
lead to the statistical result that Universe at $z\approx 2.5$ contained
at least 5 times the neutral gas that exists at the $z=0$ present
(cf Wolfe 1988). Subsequent surveys reached the conclusion that
the neutral content peaked around this time at $z=2$ to 3
(Wolfe et al 1995, Lanzetta et al 1995, Storrie-Lombardi et al 1996). 
A number of other indicators point to this period as the time when 
mass was most vigorously redistributed (cf review by Briggs 1999): 
(1) the comoving number
density of quasars peaked, implying the AGN were being fed  efficiently,
(2) the onset of the formation of metal-rich ionized halos
can be seen in the CIV absorption-line statistics for these
redshifts, and (3) the
comoving star-formation-rate density appears to have peaked at this time, 
as has been discussed in several presentations at this conference.

One of the principal indicators used to measure the comoving SFR density
during the peak in the action at $z\approx 2.5 $
has been the Lyman Break galaxy population (cf. Giavalisco et al 1996).
These objects appear to have nearly the comoving density of the
present day $L^*$ galaxies, and it is interesting to ponder the relation
between them and the DLa absorbers that collectively contain so
much neutral gas.  If the HI mass of present day $L^*$ galaxies
is simply scaled up to account for the DLa evolution and the
mass is assigned to LB galaxies, then $M_{HI}$ for each
LB galaxy would be a bit more than $10^{10}$M$_{\odot}$.
The statistical cross section for DLa absorption greatly exceeds
the apparent optical size of the LB objects in HST images. If the large
DLa cross section were to be assigned to LB population, then each
LB galaxy in the HST images could 
be an indicator of only the most central region of a much bigger neutral
(and metal enriched -- see Pettini in these proceedings) gas-rich system.
On the other hand, the results of CDM simulations  imply that, instead,
the large galaxies that we now classify $L^*$ built up over time
 from the coalescence of
many smaller protogalactic clumps. In this picture, the HI cross section
at $z\approx 2.5$ sensed by the DLa studies must be 
apportioned to many smaller objects,
implying that the LB objects could be representative of only
a tiny fraction of the galaxy population coexisting with the LB
galaxies.

A crucial test of galaxy evolution models will be to measure how
the neutral gas (whose baryon density appears to exceed that in stars at 
these redshifts) is distributed and how it relates to the LB galaxies.
Unfortunately, mapping the 21cm line emission from individual protogalaxies
at these redshifts will be very challenging, even with the most
optimistic design parameters for next generation radio telescopes. On the
other hand, considerable progress might be made with a straightforward
statistical method, on time scales much shorter than the 
construction of a new radio telescope. Several current generation
aperture synthesis telescopes (the Westerbork Synthesis Radio Telescope
and the Giant Metrewave Radio Telescope in India)
are equipped to observe the 21cm line redshifted to $z\approx 3$.
The field of view of these telescopes can survey several square
degrees of sky in a single integration with sufficient angular
resolution to avoid confusion among the LB galaxies. If an adequate
catalog of LB galaxies could be constructed for such a synthesis field,
with of order $10^4$ LB objects, with celestial coordinates and redshifts,
then the radio signals could be stacked, to obtain a statistical measure
of the HI content of the LB population.  This would allow the
``average HI content per LB galaxy site'' to be determined. Of course,
we would rather examine individual galaxies in order to directly
measure their physical size and perform kinematical studies to infer
the depths of their potential wells (i.e. the evolution of the
dark matter halos over time). How this might be accomplished is the
subject of the next section.

\section{Mapping gas rich galaxies at high redshift}

Considerable progress in assessing the extent and kinematics of
the DLa class of quasar absorption line system could be made
with minimal technical adaptation of existing radio facilities.
The technique requires background radio quasars or high redshift
radio galaxies with extended radio continuum emission. Some
effort  needs to be invested in surveys to find redshifted
21cm line absorption against these types of sources.  These
surveys can either key on optical spectroscopy of the quasars to
find DLa systems for subsequent inspection in the 21cm line, or
they can make blind spectral surveys in the 21cm line directly,
once the new wideband spectrometers that are being constructed for
at Westerbork and the new Green Bank Telescope are completed. Then
radio interferometers with suitable angular resolution at the redshifted
21cm line frequency must be used to map the absorption against the
extended background source. This would involve interferometer
baselines of only a few hundred kilometers -- shorter than is
typically associated with VLBI techniques, but longer than
the VLA and GMRT baselines.  The shorter  spacings in the European
VLBI Network and the MERLIN baselines
would form an excellent basis for these
experiments, although considerable effort will be required to
observe at the interference riddled frequencies outside the
protected radio astronomy bands.

\begin{figure}[h]
\putplot{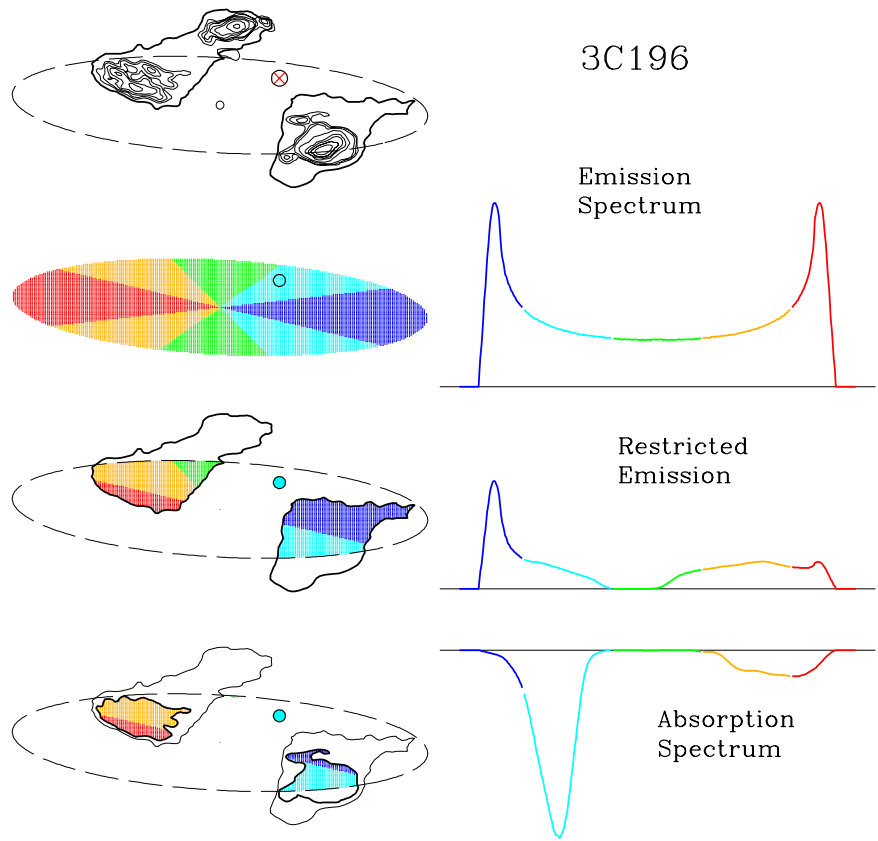}{11.0 cm}{0}{67}{67}{50}{-130}
\caption{Absorption by an intervening disk galaxy 
against an extended background radio quasar. {\it Top:}
Contours of radio continuum emission (Lonsdale 1983 with
the outer radio contour taken from 
the map of Oren as shown by Cohen et al 1996). 
{\it Upper middle:} The velocity field of and emission
profile expected for disk galaxy.
{\it Lower middle:} A spectrum that has been restricted to gas
lying in front of background continuum; in principle, sensitive
mapping could measure the distribution and kinematics for these
clouds in absorption across the face of the radio source.
{\it Bottom:} The integral absorption spectrum obtained by
observing this source with a low angular resolution telescope.
}
\end{figure}

Fig.~2 shows an example of how these experiments might work.
The top panel shows contours for the radio source 3C196. 
Brown and Mitchell (1983) discovered a 21cm line in absorption at $z=0.437$
against this source in a blind spectral survey. The object has
been the target of intensive optical and UV spectroscopy (summarized by 
Cohen et al 1996), as well as HST imaging to identify the
the intervening galaxy responsible for the absorption (Cohen et al
1996, Ridgway and Stockton 1997). Fig.~2 includes a dashed
ellipse in the top panel to indicate the approximate
extent and orientation of the galaxy identification.

The second panel from the top in Fig.~2 illustrates the 21cm line
emission spectrum typical of nearby HI-rich disk galaxies, observed
by a low resolution (``single-dish'') beam that does not resolve 
the gaseous structure in the galaxy. The rotation of a galaxy
with a flat rotation curve produces the velocity field shown
to the left of the spectrum

For disk systems observed in absorption, the information accessible to
the observer is less, since we can only hope to ever learn about the
gas opacity and kinematics for regions that fall in front of background
continuum. This restricts our knowledge to zones outlined in the
third panel of Fig.~2. The ``restricted emission'' spectrum is
drawn to illustrate what fraction of the galaxies gas content might
be sensed by a sensitive synthesis mapping observation. A comparison
to the total gas content in the upper spectrum suggests that much of
the important information (velocity spread, for example)
would be measured by a synthesis map of the absorption against
background source.

The single-dish spectrum of the absorption lines observed for an
object like 3C196 is weighted by the regions where the background
continuum has the highest brightness.  As shown in the lower panel,
this weighting emphasizes the bright spots in the radio lobes. 
Clearly sensitive mapping will better recover the information 
lost in the integral spectrum produced by a low angular
resolution observation.  A preliminary look at recent observations
of the $z=0.437$ absorber in 3C196 can be found in de Bruyn et al
(1997).

\section{Conclusion}

Neutral gas clouds rely on confinement in gravitational
potential wells in order to maintain sufficient density
that they are not ionized by the background radiation
emitted by star forming regions and AGN.  In this sense,
the damped Lyman-$\alpha$ absorbers are sign posts that
draw our attention to the evolving potential wells
of galaxies and protogalactic clumps, perhaps before they
become sufficiently luminous to be studied through their
optical or UV emission.  The 21cm line then becomes a
convenient probe of the cold gas that traces these
possibly primitive potential wells in which galaxies
form.

\section{Acknowledgements}

I am grateful to the Organizing Committee of the XIXth Recontres
de Moriond for the invitation and support for attending this
excellent conference.   


%
\vfill

\begin{moriondbib}
\bibitem{bli99}Blitz, L., Spergel, D.~N., Teuben, 
   P.~J., Hartmann, D., \& Burton, W.~B. 1999, ApJ, 514, 818
\bibitem{bra99}Braun, R., \& Burton, W.~B 1999, 
   A\&A, 341, 437 
\bibitem{bri99} Briggs, F.H. 1999, in Highly Redshifted Radio Lines, 
  ASP Conf. Series Vol. 156, Ed. by C. L. Carilli, S. J. E. Radford, 
  K. M. Menten, \& G. I. Langston, p. 16
\bibitem{bro83} Brown, R.L., \& Mitchell, K.J. 1983, ApJ, 264, 87 
\bibitem{bur96} Burles, S., \& Tytler, D. 1996, ApJ, 460, 584
\bibitem{coh96} Cohen, R.D., Beaver, E.A., Diplas, A., Junkkarinen, V.T.,
  Barlow, T.A., \& Lyons, R.W. 1996, ApJ, 456, 132            
\bibitem{deb97} de Bruyn, A.G., Briggs, F.H., \& Vermeulen, R.C. 1997,
 http://www.nfra.nl/nfra /newsletter/1997-1/index.htm
\bibitem{fan99} Fan {\it et al} 1999, AJ, 118, 1 
\bibitem{gia96} Giavalisco, M., Steidel, C.C., \& Macchetto, F.D. 1996, ApJ,
  470, 189 
\bibitem{hen95} Henning, P.A. 1995, ApJ, 450, 578
\bibitem{hen98} Henning, P.~A., Kraan-Korteweg, 
   R.~C., Rivers, A.~J., Loan, A.~J., Lahav, O., \& Burton, W.~B
   1998, AJ, 115, 584
\bibitem{kil99} Kilborn, V., Webster, R.~L., \& Staveley-Smith, L. 
   1999, PASA, 16, 8
\bibitem{lan99} Klypin, A., Kravtsov, A.V., Valenzuela, O., \& Prada, F. 1999,
   ApJ, 522, 82
\bibitem{kra99}Kraan-Korteweg, R.~C., 
   van Driel, W., Briggs, F.~H., Binggeli, B., \& Mostefaoui, T. I. 
   1999, A\&AS, 135, 255
\bibitem{lan95} Lanzetta, K.L., Wolfe, A.M., \& Turnshek, D.A. 
   1995, ApJ, 440, 435 
\bibitem{lon83} Lonsdale, C.J., \& Morison, I. 1983, MNRAS, 203, 833 
\bibitem{mat98}Mateo, M.~L. 1998, ARAA, 36, 435
\bibitem{moo99}Moore, B., Ghigna, S., Governato, F., Lake, G.,
  Quinn, T., Stadel, J., \& Tozzi, P. 1999, ApJL, 524, L19
\bibitem{put98}Putman, M.~E.  et al. 1998, Nature, 394, 752
\bibitem{rau97} Rauch, M., Miralda-Escude, J., Sargent, W.L. W.,
  Barlow, T.A., Weinberg, D. H., Hernquist, L., Katz, N.,
  Cen, R., \& Ostriker, J.P. 1997, ApJ, 489, 7
\bibitem Ridgway, S.E., \& Stockton, A. 1997, AJ, 114, 511 
\bibitem{sch91} Schneider, D.P., Schmidt, M., \&
  Gunn, J.E. 1991, AJ, 102, 837  
\bibitem{shu96} Shull, J.M., Stocke, J.T., \& Penton, S. 1996, AJ, 111, 72
\bibitem{sor94} Sorar, E.  1994, Ph.D.  Thesis, University 
   of Pittsburgh
\bibitem{spi99} Spitzak, J.~G., \& Schneider, S.~S. 1999, ApJS, 119, 159 
\bibitem{sto96} Storrie-Lombardi, L. J., McMahon, R. G., and Irwin, M. J.
   1996, MNRAS, 283, L79
\bibitem{szo94}Szomoru, A., Guhathakurta, P., van 
   Gorkom, J.~H., Knapen, J.~H., Weinberg, \& D.~H., Fruchter, A.~S. 1994,
   AJ, 108, 491
\bibitem{tos98} Tosi, M., Steigman, G., Matteucci, F., \& Chiappini, C.
  1998, ApJ, 498, 226
\bibitem{tyt96} Tytler, D., Fan, X.-M., \& Burles, S. 1996, Nature, 381, 207
\bibitem{alb85} van Albada, T.S., Bahcall, J.N., Begeman, K., \& Sancisi, R.
   1985, ApJ, 295, 305
\bibitem{wak91}Wakker, B.~P., \& van Woerden, H. 1991, A\&A, 250, 509
\bibitem{wak97}Wakker, B.~P., \& van Woerden, H. 1997, ARAA, 35, 217
\bibitem{wei91}Weinberg, D.~H., Szomoru, A., 
   Guhathakurta, P., \& van Gorkom, J.~H.  1991, ApJ, 372, L13
\bibitem{wol88} Wolfe, A.M. 1988, in QSO 
    Absorption Lines: Probing the Universe, eds. Blades, J.C.
    Turnshek, D.A., \& Norman, C.A. 1988, p. 297
\bibitem{wol95} Wolfe, A.M., Lanzetta, K.M., Foltz, C.B., Chaffee, F.H.
  1995, ApJ, 454, 698
\bibitem{zwa97} Zwaan, M.A., Briggs. F.H., Sprayberry, D., \& Sorar, E.
  1997, ApJ, 490 173
\bibitem{zwa99} Zwaan, M.A., \& Briggs. F.H  1999, ApJLett, in press,
astro-ph/0001016
\end{moriondbib}
\end{document}